# A Key Pre-Distribution Scheme based on Multiple Block Codes for Wireless Sensor Networks

Hamidreza Arjmandi, Farshad Lahouti
Center for Wireless Multimedia Communications
School of Electrical & Computer Engineering
University of Tehran, Tehran, Iran
Emails: h.arjmandi@alumni.ut.ac.ir, lahouti@ut.ac.ir

*Abstract*—A key pre-distribution scheme (KPS) based on multiple codewords of block codes is presented for wireless sensor networks. The connectivity and security of the proposed KPS, quantified in terms of probabilities of sharing common keys for communications of pairs of nodes and their resilience against colluding nodes, are analytically assessed. The analysis is applicable to both linear and nonlinear codes and is simplified in the case of maximum distance separable codes. It is shown that the multiplicity of codes significantly enhances the security and connectivity of KPS at the cost of a modest increase of the nodes storage. Numerical and simulation results are provided, which sheds light on the effect of system parameters of the proposed KPS on its complexity and performance. Specifically, it is shown that the probability of resilience of secure pairs against collusion of other nodes only reduces slowly as the number of colluding nodes increase.

## I. Introduction

Secure communication between nodes in WSNs involves interesting technical and research challenges[1]-[4]. The security requirements for WSNs are authentication, integrity and confidentiality. In addition, the specific features of sensor networks impose certain design requirements including key connectivity, resilience against colluding nodes, limited complexity (storage, processing and communications), and scalability[1]. In general, the security solutions for such networks relyon strong and efficient key distribution mechanisms. Considering the complexity constraints, key management in WSNs is primarily set up based on a symmetric key pre-distribution scheme (KPS). Using KPS, a special entity or authority acts as the key distribution center and loads the keys into the nodes prior to deployment [1].

For WSNs, the deterministic KPSs proposed in [5]-[7] use a multi-dimensional grid. Specifically, each node is assigned a $k$-tuple ID, which casts the nodes in a $k$-dimensional grid. These schemes guarantee that any pair of nodes can find a common key either directly or indirectly through intermediate nodes. In fact, establishing indirect keys incurs communication overheads for the intermediate nodes and relies on their trustworthiness.

In[8], a symmetric key generation schemebased on key pre-distribution is presented that uses a maximum distance separable (MDS) code to secure pairwise communication of nodes in a network. This scheme guarantees direct common key between all pairs of nodes, at the cost of a complex key discovery phase of quadratic complexity. An alternative and more efficient deterministic KPS scheme based on MDS codes is suggested in[9], which also shows a stronger resilience against collusion.In this scheme,the ID of a node is encoded by an MDS code to generate the index of keys associated with that node. In [14], a key management system based on MDS codes has been presented in which a publicly known MDS generator matrix is available to every node. Every node in the network uses this matrix and a random vector to generate codeword which is the secret key chain. In [10],a KPS using Reed-Solomon codes is suggested. For this special case, the sharing probability (probability that two nodes find direct common key) and a bound on resilience against collusion are computed. The same performance measures are assessed for linear codes in [13]. In [15],a general class of designs, termed "partially balanced $t$-designs" is introduced, that encompasses almost all of the reported designs for combinatorial key pre-distribution schemes.

In this paper, we present a key pre-distribution scheme based on multiple codewords of a block code and analyze its performance. The multiplicity of the codewords facilitates stronger security and may be administered by a single or multiple key assignment authorities. As presented in the next Section, the proposed scheme and the analyses of its sharing and resilience probabilities are applicable to all types of linear or nonlinear codes. Specifically, we derive an exact analytical expression to quantify the probability of resilience of the proposed KPS against $r$ colluding nodes thatparticularly simplifies in the special case of MDS codes. The results show that the resilience probability improves exponentiallyfast with the code multiplicity at the cost of a modest increase of storage requirements.Numerical and simulation results in Section III demonstrate the accuracy of the analysis and quantify the performance and complexity of the proposed KPS. The article is concluded in Section IV.

## II. MBKPS and Resilience Analysis

A block code $C(n,k)-q$ over a field of $q$ symbols, $F_q$, is a set of $q^k$ vectors of length $n$ called codewords. Associated with this, is an encoder which maps an input$k$-tuple over $F_q$ to its corresponding codeword [12]. Below, we present a key pre-distribution scheme based on multiple codewords of such a block code, herein referred to as Multiple Block Code Key Pre-Distribution Scheme, MBKPS. For better presentation, we begin with the case where only one codeword is used for key assignment of each single node (BKPS).

### A. Key assignment in BKPS

Consider a network with $N$ nodes. In key assignment phase of the scheme, we first assign each node an ID as a $q$-ary vector of $k$ symbols where $N \leq q^k$. Then this ID is encoded by a block code $C(n,k)$-$q$ producing a codeword which serves as key-index ID of the node. As described in Table 1, corresponding to each symbol of key-index ID of a node, one key is assigned to this node.

Table 1. Key assignment algorithm of BKPS

1. A $q$-ary vector of length $k = \lceil \log_q N \rceil$ is assigned to each node as its ID.
2. For node $A$, its ID $\widehat{\alpha_k} = \{\hat{\alpha}_0, \hat{\alpha}_1, \ldots, \hat{\alpha}_{k-1}\}$ is encoded by the block code $C(n,k)$-$q$ to produce its key-index ID $\boldsymbol{\alpha}_{(n,k)} = \{\alpha_0, \alpha_1, \ldots, \alpha_{n-1}\}, \alpha_i \in F_q, \forall i$.
3. Prepare a key pool with $q \times n$ keys $\boldsymbol{K} = \{k_0, \ldots, k_{qn-1}\}$.
4. The keys $k(i)$, $0 \leq i \leq n-1$, is assigned to node A with key-index ID, $\alpha_{(n,k)}$, as follows
$$k(i) = k_{f(i)}, \quad f(i) = \alpha_i \times n + i, \quad k_{f(i)} \in \boldsymbol{K}$$

It is noteworthy that the presented BKPS is technically equivalent to the KPS proposed in [10], however, the key assignment formulation presented here is more concise. Specifically, the BKPS algorithm in Table 1 relies on a mapping function, $f(.)$, in step 4 that is a function of a single variable, whereas the equivalent mapping in the KPS of [10] is a function of two variables.

### B. Key assignment in mutiple BKPS

To enhance the security and/or reducing the dependency on trust to one authority, we propose the multiple BKPS (MBKPS). For a scenario with $M$ authorities, each node is assigned one ID of length $k$ by each of the authorities independently. Each $k$ symbol ID is encoded by the block code $C(n,k)$-$q$. Therefore, a $q$-ary key-index ID of length $M \times n$ is assigned to each node. Each authority assigns the keys to its associated $n$ symbols in the key-index ID. The MBKPS is secure against collusion of less than $M$ authorities as it does not allow revealing all common keys between the two nodes. Table 2 shows the key assignment algorithm of the MBKPS.

Table 2. Key assignment algorithm of MBKPS

1. An ID is assigned to each node as a vector of length $k$ over $F_q$ by each authority $m$, $0 \leq m \leq M-1$ where $k = \lceil \log_q N \rceil$ and $N$ is the number of nodes.
2. For node $A$, its ID $\{\hat{\alpha}_{mk}, \hat{\alpha}_{mk+1}, \ldots, \hat{\alpha}_{(m+1)k-1}\}$ (assigned by authority $m$) is encoded by the block code $C(n,k)$-$q$ to produce its associated part of key-index ID $\{\alpha_{mn}, \alpha_{mn+1}, \ldots, \alpha_{(m+1)n-1}\}, \alpha_i, \hat{\alpha}_i \in F_q, \forall i$.
3. The authority $m$, $0 \leq m \leq M-1$ prepares a key pool containing $n \times q$ keys $\boldsymbol{K}_m = \{k_{mn}, \ldots, k_{(m+1)nq-1}\}$.
4. The authority m assigns keys $k(i)$, $mn \leq i \leq (m+1)n-1$, to node A based on its key-index ID as follows:
$$k(i) = k_{f(i)}, \quad f(i) = \alpha_i M n + i, \quad k_{f(i)} \in K_m$$

For $M = 1$, the presented MBKPS reduces to BKPS.

**Remark 1.** The MBKPS is presented assuming $M$ authorities. Alternatively, a single authority may opt for MBKPS for enhanced security. This is accomplished when it assigns $M$ independent key-index IDs of length $n$ to a node and runs the BKPS for every part with independent key pools.

**Property 1.** Using $C(n,k)$-$q$ and based on MBKPS key assignment, $M \times n$ distinct keys are assigned to each node.

**Property 2.** The proposed MB KPS with $M$ authorities and using the code $C(n,k)$-$q$ incurs the local and global costs of $M \times n$ and $M \times n \times q$, respectively. This indicates that both costs increase linearly as the number of authorities increases.

**Property 3.** Each authority assigns its own portion of the node ID and key-index ID independent of other authorities. Hence, if a collusion of authorities/nodes has access to the common keys of two nodes due to a given part of their key-index IDs, they may still be protected against collusion by keys corresponding to other parts of their key-index IDs. As a result, if the probability that all common keys related to one authority are identified by a collusion of nodes is $P$, this probability decreases to $P^M$ by $M$ authorities.

The key discovery phase of the MBKPS is very simple and of low complexity. To find common keys, two nodes exchange their key-index IDs. Each node then finds common keys by comparing the two key-index IDs and identifying the common symbols. The probability that two nodes find at least one common key is defined as sharing probability. Based on the computed sharing probability of BKPS in [10], we provide the sharing probability of MBKPS as follows.

**Proposition 1.** Using blockcode $C(n,k)$-$q$, if $P_{sh}$ is the sharing probability of BKPS, for MBKPS key assignment with $M$ authorities, the sharing probability is:
$$P_{sh}^M = 1 - (1 - P_{sh})^M. \tag{1}$$

**Proof:** The probability that two nodes do not find any common key associated with one authority is $1 - P_{sh}$. As $M$ key-index ID parts are assigned independently, then the probability that two nodes do not have any common key in $M$ parts of their key-index ID is equal to $(1 - P_{sh})^M$. Hence, the sharing probability of MBKPS for $M$ authorities is obtained as in (1).

As evident from Proposition 1, the sharing probability approaches 1 as $M$ increases.

### C. $r$-resilience of MBKPS

The resilience of a key pre-distribution scheme against colluding nodes, $r$-resilience, is defined as the ratio of the secure pairwise communication channels over the total number of node pairs. A pair of nodes is secure against collusion of $r$ nodes, if there exists a common key between these two nodes that is not in the union of the set of keys assigned to the $r$ colluding nodes. The corresponding symbol with such a secure key is referred to as a collusion free symbol.

**Proposition 2.** Using block code $C(n,k)$-$q$ for BKPS, the total number of node pairs that has secure common key against an $r$-collusion is

$$D = \sum_{j=1}^{n-d_{min}} \sum_{i_1,i_2,\ldots,i_j \in \{1,\ldots,n\}} (-1)^{j-1} V_{i_1,i_2,\ldots,i_j} \quad (2)$$

$$V_{i_1,i_2,\ldots,i_j} = \sum_{\substack{q_1,q_2,\ldots,q_j \\ q_t \in U_{i_t}, \forall 1 \le t \le j}} \binom{H_{i_1,i_2,\ldots,i_j}^{q_1,q_2,\ldots,q_j}}{2} \quad (3)$$

in which, $H_{i_1,i_2,\ldots,i_j}^{q_1,q_2,\ldots,q_j}$ is the number of codewords with symbol values $q_1, q_2, \ldots, q_j$ in coordinates $i_1, i_2, \ldots, i_j$, respectively and $U_{i_t}$ is the set of collusion free symbols in coordinate $i_t$.

**Proof:** Each pair of the nodes is secure against $r$-collusion, if they have identical collusion free symbols in the same positions in their key-index IDs. Consider $\mathcal{D}$ as the set of all pairs of nodes secured against $r$-collusion. Let $D$ represent the number of elements of the set $\mathcal{D}$. It is clear that $\mathcal{D}$ can be viewed as the union of the sets $\mathcal{V}_i$, $i = 1, \ldots, n$, in which $\mathcal{V}_i$ is the set of all pairs of nodes with secure key (common collusion free symbol) in coordinate $i$, i.e., $\mathcal{D} = \bigcup_{i=1}^{n} \mathcal{V}_i$. Let $V_i$ and $V_{i_1,i_2,\ldots,i_j}$ denote the number of elements of the sets $\mathcal{V}_i$ and $\bigcap_{t=1}^{j} \mathcal{V}_{i_t}$, $i_t \in \{1, \ldots, n\}$. Therefore, the number of pairs that have at least $j$ common collusion free symbol symbols is equal to $\sum_{i_1,i_2,\ldots,i_j \in \{1,\ldots,n\}} V_{i_1,i_2,\ldots,i_j}$. Hence, the number of pairs having at least one common symbol, $D$, is given by (2). Note that the term $(-1)^{j-1}$ in (2) help deduct the pairs with more than $j$ common symbols in $\sum_{i_1,i_2,\ldots,i_j \in \{1,\ldots,n\}} V_{i_1,i_2,\ldots,i_j}$. As the hamming distance of the code is $d_{min}$, a pair of nodes have at most $n - d_{min}$ identical symbols. As a result, for $j > n - d_{min}$, we have $V_{i_1,i_2,\ldots,i_j} = 0$.

Now we compute $V_{i_1,i_2,\ldots,i_j}$ as follows. $V_{i_1,i_2,\ldots,i_j}$ is the number of pairs of nodes with identical collusion free symbols in all coordinates $i_1, i_2, \ldots, i_j$. Therefore, it can be rewritten as follows

$$V_{i_1,i_2,\ldots,i_j} = \sum_{\substack{q_1,q_2,\ldots,q_j \\ q_t \in U_{i_t}, \forall 1 \le t \le j}} V_{i_1,i_2,\ldots,i_j}^{q_1,q_2,\ldots,q_j} \quad (4)$$

in which, $V_{i_1,i_2,\ldots,i_j}^{q_1,q_2,\ldots,q_j}$ is the number of pairs of nodes with identical collusion free symbols of $q_1, q_2, \ldots, q_j$ in coordinates $i_1, i_2, \ldots, i_j$. Consider $H_{i_1,i_2,\ldots,i_j}^{q_1,q_2,\ldots,q_j}$ as the number of codewords with symbol values $q_1, q_2, \ldots, q_j$ in coordinates $i_1, i_2, \ldots, i_j$.

Obviously, we have $V_{i_1,i_2,\ldots,i_j}^{q_1,q_2,\ldots,q_j} = \binom{H_{i_1,i_2,\ldots,i_j}^{q_1,q_2,\ldots,q_j}}{2}$, and the proof is complete. ∎

The $r$-resilience probability of the BKPS using block code $C(n,k)$-$q$ is represented by $P_{re}$ and is defined as the ratio of $D$ over the total number of choices of pairs of nodes, i.e,

$$P_{re} = \frac{D}{\binom{q^k}{2}} \quad (5)$$

in which $D$ is given by Proposition 2.

**Remark 2.** The proposed analysis of the $r$-resilience probability of BKPS holds generally for any type of block code and is not restricted to linear codes.

**Proposition 3.** Using block code $C(n,k)$-$q$ for MBKPS key assignment with $M$ authorities, the $r$-resilience probability is

$$P_{re}^M = 1 - (1 - P_{re})^M. \quad (6)$$

**Proof:** The probability that two nodes do not find any common collusion free symbol (secure key) associated with one authority is $1 - P_{re}$. As $M$ key-index ID parts are assigned independently, then the probability that two nodes do not have any common secure key in $M$ parts of their key-index ID is equal to $(1 - P_{re})^M$. Hence, the $r$-resilience probability of MBKPS for $M$ authorities is obtained as in (6). As evident in Proposition 3, the $r$-resilience probability approaches 1 as $M$ increases.

### D. $r$-resilience of MBKPS for MDS codes

For a linear code $C(n,k) - q$, the number of codewords with the same symbol in a coordinate is exactly $q^{k-1}$ [11]. The value of each coordinate is distributed uniform over $GF(q)$ for linear codes over this field. Hence, the probability that none of the $r$ colluding nodes at a given coordinate assumes a certain symbol is $(1 - \frac{1}{q})^r$. As such, the average number of collusion free symbols in a given coordinate for a linear code is given by $u = q(1 - \frac{1}{q})^r$.

For MDS codes, the number of codewords with identical symbols in $1 \le t \le k$ positions is exactly $q^{k-t}$ [11]. Therefore, for the MDS code $C(n,k)$-$q$, $d_{min} = n - k + 1$, and we have:

$$H_{i_1,i_2,\ldots,i_j}^{q_1,q_2,\ldots,q_j} = q^{k-j}, \ 0 \le j \le k - 1. \quad (7)$$

It is straight forward to obtain the following

$$D = \sum_{j=1}^{k-1} (-1)^{j-1} u^j \binom{n}{j} \binom{q^{k-j}}{2}. \quad (8)$$

## III. PERFROMANCE EVALUATION

In this Section, the performance of the proposed BKPS is assessed via numerical and simulation results. We consider a network with $N$ nodes of limited storage to store the codewords and keys. In order to establish MBKPS based on $C(n,k)$-$q$ with $M$ authorities, each node needs to store $M \times n$ symbols related to its key-index ID and also $M \times n$ keys. If the keys are chosen from the field $F_{q'}$, then each node needs $S = Mn \log_2 q + Mn \log_2 q'$ bits of storage.

Fig.1 depicts the simulationresults for the $r$-resilience of BKPS for two MDS codes of (14,2)-16 and (30, 2)-32. For comparison, we consider the average performance of fifty randomly generated linear codes with the same parameters. The results are averaged over random sets of colluding nodes. It is observed that for BKPS with thesamecode parameters, $n$, $k$, $q$, the MDS code demonstrates a strongerresilience.In addition, the figuredemonstrates that the simulations coincide with the proposed analysisof BKPS resilience performance.

Fig. 2demonstrates how increasing the memory storage of nodes can be used by the proposed MBKPS algorithm forimprovingthe resilience of the system.The $r$-resilience is evaluated for $N = 1000$, $q' = 64$ and certaincodes andvalues of $M$. It is observed that a largernode storage, $S$ (bits),when effectively used within the MBKPSimproves the resilience of the system significantly. For example increasing $S$ from 20 kbits to 50kbits can improve resilience against 100 colluding nodes by about 100%.

Fig.3 compares the performance of the proposed analysis of resilience (which coincides with the true performance) and the lower bound proposed in [10] based on the definition in Section II.C. It is evident that the performance gap increases with$r$, and affirms the value of the proposed analysis.

## IV. CONCLUSIONS

In this paper, a key pre-distribution scheme based on multiple block codes was proposed for WSN applications. While this multiplicity may be administered by single or multiple key assignment authorities, it was analytically proven that it improves the sharing and resilience probabilities exponentially fast, at the cost of only a linear increment of memory storage. The presented analysis also demonstrated that the resilience probability approaches zero exponentially slow as the number of colluding nodes grows. Future research in this direction could investigate the optimal design of codes and system parameters under specific complexity constraints for nodes.

## REFERENCES

[1] M.A. Simplício Jr. et al., "A survey on key management mechanisms for distributed wireless sensor networks," *Computer Network,* April2010.

[2] Y. Xiao, V. Rayi, B. Sun, X. Du, F. Hu, M. Galloway, "A survey of key management schemes in wireless sensor networks," *Elsevier Computer Commun.,* Vol. 30, No. 11–12, pp. 2314–2341, 2007.

[3] Y. Zhou, Y. Fang, Y. Zhang, "Securing wireless sensor network," *IEEE Commun. Surveys and Tutorial,* Vol. 10, No. 3, pp. 6-28, 2008.

[4] J. Zhang and V. Varadharajan, "Wireless sensor network key management survey and taxonomy," *Elsevier Journal of Network and Computer Applications,* Vol. 33, No. 2, pp. 63–75, 2010.

[5] Y. Zhou and Y. Fang, "Scalable and deterministic key agreement for large scale networks," *IEEE Trans. Wireless Commun.,* Vol. 6, No. 11, Nov. 2007.

[6] H. Chan and A. Perrig, "PIKE: Peer intermediaries for key establishment in sensor networks," *Proc. INFOCOM,* Miami, FL, Mar. 2005.

[7] F. Delgosha and F. Fekri, " A multivariate key-establishment scheme for wireless sensor networks," *IEEE Trans. Wireless Commun.,* Vol. 8, No. 4,pp. 1814 – 1824, 2009.

[8] R. Blom, "Non-public key distribution," Advances in Cryptology, 1982.

[9] Y. E. Yang and J. D. Touch, "Protocol family for optimal and deterministic symmetric key assignment," *IEEE Int. Conf. Networking,* 2008.

[10] S. Ruj, B. Roy, "Key predistribution schemes using codes in wireless sensor networks,"*Information Security and Cryptography, Lecture Notes in Computer Science,* Vol. 5487, pp. 275-288, 2009.

[11] T. Richardson, R. Urbanke,Modern coding theory, Cambridge University Press, 2008.

[12] F. J. MacWilliams and N. J. A. Sloane, The theory of error-correcting codes, 1977.

[13] Q. Chen, D. Pei, J. Dong, "Determining parameters of key predistribution schemes via linear codes in wireless sensor networks," *Information Security and Cryptography, Lecture Notes in Computer Science,*Vol. 6584, pp. 284-299, 2011.

[14] M. Al-Shurman, S.M.Yoo, "Key pre-distribution using MDS codes in mobile ad hoc networks,"*Int. Conf. Information Technology: New Generations,* pp. 566–567, 2006.

[15] M. B. Paterson, D. R. Stinson, "A unifiedapproach to combinatorial key pre-distribution schemesfor sensor networks" *Designs, Codes and Cryptography*, No. 3, Vol. 71, June 2014.

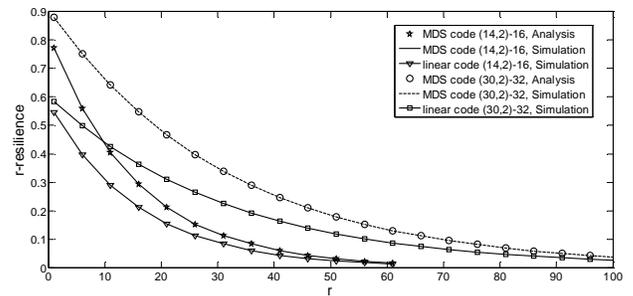

Fig.1.$r$-resilience vs. number of colluding nodes forBKPSwith different MDS and random linear codes.

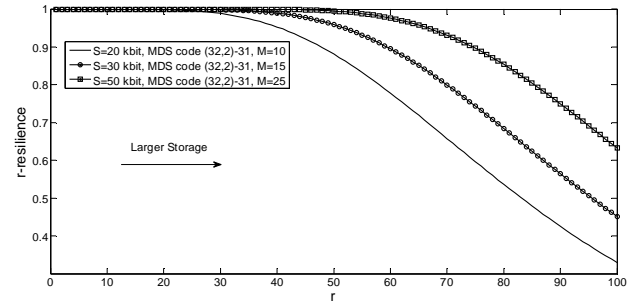

Fig.2. $r$ -resilience vs. number of colluding nodes forMBKPSwith MDS codes and different values of $S$; $N = 1000$.

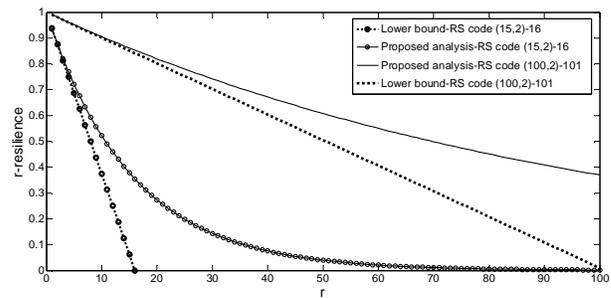

Fig.3. Comparison of the proposed $r$-resilience probability with the $r$-resilience lower bound provided in[10].